\documentclass[preprint, aip,jcp]{revtex4-1}

\usepackage{graphicx} 
\usepackage{color}

\begin{document}

 \title{
An observable for vacancy characterization and diffusion in crystals.
}

\author{Pierre-Antoine Geslin}
\altaffiliation{Laboratoire d'Etudes des Microstructures (LEM), CNRS-ONERA, 29 Avenue de la Division Leclerc, BP 72, 92322 Chatillon, France.}
\affiliation{
Ecole Nationale Sup\'erieure des Mines de Sant-\'Etienne, 158 Cours Fauriel, 42023 Saint-\'Etienne cedex 2, France.}

\author{Giovanni Ciccotti}
\altaffiliation{Dipartimento di Fisica and CNISM, Universit\`a ``La Sapienza'', Piazzale Aldo Moro 5, 00185 Rome, Italy.} 
\affiliation{School of Physics, University College Dublin, Belfield, Dublin 4, Ireland.}

\author{Eric Vanden Eijnden}
\affiliation{Courant Institute of Mathematical Sciences, New York University, New York, New York 10012, USA.}

\author{Simone Meloni}
\altaffiliation{Permanent address: Consorzio Interuniversitario per le Applicazioni di Supercalcolo Per Universit\`a e Ricerca (CASPUR), Via dei Tizii 6, 00185 Roma, Italy.}
\email{To whom correspondence should be addressed: s.meloni@cineca.it}
\affiliation{School of Physics, University College Dublin, Belfield, Dublin 4, Ireland.}

 \date{\today}

\begin{abstract}
\noindent \normalsize{To locate the position and characterize the dynamics of a vacancy in a crystal, we propose to represent it by the ground state density of a quantum probe quasi-particle for the Hamiltonian associated to the potential energy field generated by the atoms in the sample. In this description, the $\hbar^2/2 \mu$ coefficient of the kinetic energy term is a tunable parameter  controlling the density localization in the regions of relevant minima of the potential energy field.
Based on this description, we derive a set of collective variables that we use in rare event simulations to identify some of the vacancy diffusion paths in a 2D crystal. Our simulations reveal, in addition to the simple and expected nearest neighbor hopping path, a collective migration mechanism of the vacancy. This mechanism involves several lattice sites and produces a long range migration of the vacancy. Finally, we also observed a vacancy induced crystal reorientation process.}
\end{abstract}

\maketitle

\section{Introduction}

Far from the melting, diffusion in solids arise when atoms migrate into empty sites in the crystal, leaving other empty sites behind them in which atoms can again migrate. Each of these empty sites is referred to as a vacancy, and the corresponding diffusion mechanism is said vacancy-driven. The atom/vacancy diffusion is a thermally activated process requiring the system to overcome free energy barriers separating the initial from the final state. In most cases, these free energy barriers are much higher than the thermal energy, therefore this process is a rare event, i.e. an event occurring with a frequency that is too low  to be sampled by ``brute force'' Molecular Dynamics (MD) or Monte Carlo (MC) simulations. Its study requires special simulation techniques, such as  
{Temperature Accelerated MD(TAMD)/Temperature Accelerated MC (TAMC),\cite{TAMD,TAMC} Metadynamics~\cite{Metadynamics, Metadynamics2} or Adiabatic Free Energy Dynamics (AFED)~\cite{AFED} {to efficiently explore the free energy surface of the system, or the string method in collective variables~\cite{Maragliano2006b} to identify statistically relevant paths.} These approaches require suitable collective variables (CVs) to describe the rare event. 

{If one is not interested in the free energy, other methods exist that do not require CVs, or for which the identification of good CVs is not crucial. These methods can be classified in two groups: methods to identify the transition mechanism in the configuration space  (e.g.  the Nudged Elastic Band,\cite{Jonsson1998,Henkelman2000} the dimer method,\cite{Henkelman1999} the Transition Path Sampling~\cite{1998JChPh.108.1964D}), and methods for exploring potential energy surfaces, searching for their mechanical equilibrium points and saddle points (e.g. the Temperature Accelerated Dynamics,\cite{Sorensen2000} the Hyperdynamics,\cite{Voter1997} and the ART method.\cite{Barkema1996,Mousseau}) However, also these methods suffer from limitations (in addition to that already mentioned of not allowing to reconstruct the free energy). As for the first group, they require the {\itshape a priori} knowledge of the initial and final states of the process, and a tentative path belonging to the ``reactive channel'' one is interested in. As for the second group, despite significant progresses,\cite{Barkema1996,Mousseau} typically their reliability degrades with the complexity of the system. It must be stressed that these methods have been widely used to study defects (vacancy, interstitial, etc.) migration  processes,\cite{Song:2001em,Gao:2003jw,Zobelli:2007eq} especially in ``simple'' systems. It must be also stressed that having a simplified but more expressive description of the vacancy migration process, like that attainable by CVs, is desirable as it allows to provide a more thoroughly physical picture of the events of this type, for example by effectively representing cooperative or collective effects.}

The identification of good CVs for modeling the vacancy migration is still only roughly solved even for the most simple mechanism: the ``local'' vacancy migration mechanism. This mechanism consists in the hopping of the vacancy into a nearest neighbor lattice site. In this simple case, CVs have been proposed, e.g. in Refs.~\onlinecite{Bennet}, \onlinecite{1987JPhC...20.2331G} and \onlinecite{PaciCiccotti}, but turned out to be not completely satisfactory.\cite{PaciCiccotti}
Moreover, more complex mechanisms have been suggested to take place in Bravais lattices and lattices with basis. For example, Da Fano and Jacucci found that the modeling of high temperature diffusion in Al and Na requires the inclusion of the vacancy double jump mechanism.\cite{DaFanoJacucci} Another case of complex vacancy migration mechanism is the collective migration of proton vacancies in hydrogen bonded network materials. For example, it was found that the collective proton transfer in [dabcoH]$^+$[ReO$_4$]$^-$ (dabco = diazabicyclo[2.2.2]octane) is at the basis of the formation of the ferroelectric phase of this material.\cite{hydrogenBond1} {Vacancy migration in this material is coupled with the dynamics of the molecules forming the hydrogen bond network. Thus, a suitable CV must be able to take into account this phenomenon.}  In even more complex crystals, such as clathrate hydrates of molecular gases, the vacancy is associated to (missing) guest molecules (CH$_4$, CO$_2$, etc.) and their diffusion mechanism requires the cooperative effect of the water molecules forming the framework of the crystal.\cite{Trout} {Also in this case, the CV should be able to take into account the concerted dynamics of the framework and guest molecules.} 
In materials with a high concentration of vacancies, such as yttria and scandia stabilized cubic zirconia, vacancy-vacancy correlations were found to play a crucial role in the mass transport mechanism (for example of oxygen in the case of zirconia~\cite{Pietrucci})and using methods based on the configuration space it might result difficult to characterize this phenomenon.\cite{Note1}
Finally, the vacancy diffusion is proposed to be the limiting step in the nucleation and growth of some nanostructures, for example in Cu$_2$S nanowires.\cite{Liu2010} {Studying the mechanism of this process requires a description of the vacancy, and its migration, that is independent on the orientation of the growing nucleus (which is not known in advance), robust with respect to some degree of disorder that might be present in small crystal-like nuclei, or even to the change of crystalline symmetry along its growth.}

While the description of the simple local migration mechanism in Bravais lattices can be obtained by some {\itshape ad hoc} CVs, the treatment of the more complex cases mentioned above requires the introduction of ``general'' CVs to represent  vacancies and their dynamics. By ``general'' we mean CVs that are not specific to any crystal symmetry or orientation, or tailored to monitor a specific migration path. {Another requirement is that the CVs work also with ``stressed'' crystalline systems, such as nano crystals, that might by non-uniform or non isotropic (see, for example, Refs.~\onlinecite{boninelli,kelires,Orlandini}).} The aim of this paper is to introduce such a general set of CVs. {We illustrate the use of our CVs by investigating} vacancy migration paths in a 2D crystal of purely repulsive Lennard-Jones particles (Week-Chandler-Andersen particles~\cite{WCA} - WCA). We believe that our CVs can be used also in more complex systems of the kind mentioned above.

The remainder of this paper is organized as follows. In Sec.~\ref{sec:QuantumProbeParticle} 
we introduce the general formalism of our description of the vacancy,  and derive a set of CVs that are numerically more convenient to study the vacancy migration process. In Sec.~\ref{sec:ResultsAndDiscussion} we test our approach by first characterizing the vacancy migration observed during a high temperature MD trajectory of a simple 2D WCA crystal, and then using a reduced CV set in combination with TAMC to find interesting migration paths (and other events) in the same system. In Sec.~\ref{sec:Conclusions} we draw some conclusions.

\section{The quantum probe quasi-particle representation of a vacancy.}
\label{sec:QuantumProbeParticle}
In this section we introduce the observable that we consider the most adequate to represent a vacancy and track its dynamics. We postpone to Appendix \ref{App:simplerApproaches} the description of simpler alternatives that, although more intuitive to grasp, are not completely satisfactory.

 \begin{figure}
 \begin{center}
 \includegraphics[width=0.3\textwidth]{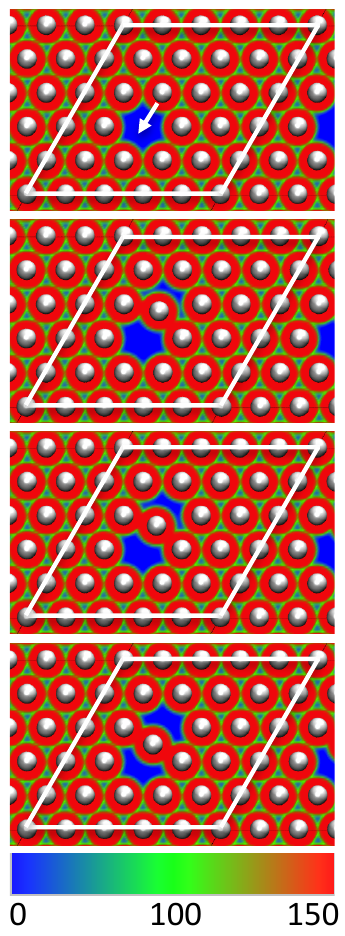}
 \end{center}
 \caption{Series of atomic configurations, and the corresponding $V({\bf x}; {\bf R})$ potential, along the local migration path in a 2D trigonal lattice of WCA particles (see Sec. \ref{sec:ResultsAndDiscussion-AnalysiOfTheDensity} for all the details). $V({\bf x}; {\bf R})$ is reported in Lennard-Jones units {(these units are used throughout the text)} and represented according to the colormap reported in the bar at the bottom of the figure. The simulation box is also reported in the figure (periodic boundary conditions are used in all calculations discussed in the text). The top panel corresponds to the configuration in which the vacancy is located on a lattice site. The arrow indicates the path followed by the moving atom during the (ideal) local vacancy migration process described in the text. In the next two panels are shown the atomic configuration and the potential $V({\bf x}; {\bf R})$ at intermediate states. In these states, the potential $V({\bf x}; {\bf R})$ presents already the characteristic ``double-well'' shape. In the bottom panel is reported the mid-way configuration and the corresponding potential $V({\bf x}; {\bf R})$. In this state, the two wells of $V({\bf x}; {\bf R})$ are symmetric.} \label{fig:FeedbackPotentialAndConfigurations}
 \end{figure}

A vacancy is the lack of an atom at a lattice site in a crystalline system. Therefore, describing a vacancy amounts to identifying the empty site in a crystal and following its evolution in time. Our starting point is to represent a vacancy as a quantum probe quasi-particle subject to the field produced by the atoms in the sample: 

\begin{equation} 
\label{eq:potential}
V({\bf x}; {\bf R}) = \sum_{i = 1}^N v(|{\bf R}_i - {\bf x}|)
\end{equation} 

\noindent where ${\bf x}$ is the position of the probe particle, ${\bf R}$ is the 3N vector of the atomic positions (${\bf R}_i$ is the position of the atom $i$) and $v(r)$ is the pair potential governing the interaction of the atoms in the system. The extension to more complex interaction potentials, including {\itshape ab initio} ones, is conceivably feasible (see Appendix~\ref{App:QuantumPotentialCase}). 
The potential energy in Eq.~\ref{eq:potential} is that of a probe particle located at ${\bf x}$ when the atoms are in the configuration ${\bf R}$. 
To give to the reader an intuitive argument of why the above potential is a key element in the vacancy description, in   Fig.~\ref{fig:FeedbackPotentialAndConfigurations} we report the $V({\bf x}; {\bf R})$ along the (ideal) local migration process in a 2D WCA trigonal crystal (more details on the calculation of $V({\bf x}; {\bf R})$ are given in Sec.~\ref{sec:ResultsAndDiscussion}). {In this process one atom, nearest neighbor of the vacancy, moves along the linear path connecting its (initial) lattice and the vacancy site.} At the beginning (top panel), the potential is characterized by one deep minimum in correspondence of the vacancy, and other local minima located in between the atoms.  
From now on, we will refer to the former as vacancy minimum (minima) and to the latter as crystal minima, as they are present also in perfect crystals.\cite{Note2}
While the atom moves toward the vacancy (the two central panels), a second vacancy minimum is formed at its original site. When the atom is mid-way along the path (bottom panel), the potential $V({\bf x}; {\bf R})$ presents two vacancy minima of equal value in the region of the two lattice sites involved in the process.

 {One could think of using directly the potential $V({\bf x}; {\bf R})$ to describe the vacancy. However, this representation is way too complex, since $V({\bf x}; {\bf R})$ is typically characterized by very many minima of which only one or few of them are relevant for the description of the vacancy migration process.} 
We propose to describe the vacancy by the ground state probability density of the quantum probe quasi-particle, $\rho({\bf x}| {\bf R}) = |\psi({\bf x}; {\bf R})|^2$, where $\psi({\bf x}; {\bf R})$ is the function minimizing the Ritz functional

\begin{equation}
\label{eq:RitzFunctional}
\psi({\bf x}; {\bf R}) = \arg \min_{ \psi} {\langle\psi({\bf x}; {\bf R}) | {\mathcal H}({\bf x}; {\bf R}) | \psi({\bf x}; {\bf R})\rangle \over  \langle\psi({\bf x}; {\bf R})|\psi({\bf x}; {\bf R})\rangle}
\end{equation}

\noindent In Eq.~[\ref{eq:RitzFunctional}] ${\mathcal H}({\bf x}; {\bf R}) = -a \nabla^2 + V({\bf x}; {\bf R})$, and $a >  0$ is a tunable parameter we will discuss shortly.\cite{Note3}
$\psi({\bf x}; {\bf R})$ can, equivalently, be defined as the ground state of the Schr\"odinger equation associated to the Hamiltomian ${\mathcal H}({\bf x}; {\bf R})$, but the formulation in terms of the Ritz functional makes more simple the analysis of some of its properties. 

In the following we will show that, for a suitable value of $a$ (discussed in detail in Sec.~\ref{sec:ResultsAndDiscussion-AnalysiOfTheDensity}), $\rho({\bf x}| {\bf R})$ is localized around the vacancy minima of the $V({\bf x}; {\bf R})$ potential and is smooth. To understand why $\rho({\bf x}| {\bf R})$ is localized around vacancy minima, consider separately the potential and kinetic energy contributions to the expectation value of the operator ${\mathcal H}({\bf x}; {\bf R})$. The potential energy part can be recast into the form $\langle \psi({\bf x}; {\bf R}) | {V}({\bf x}; {\bf R}) | \psi({\bf x}; {\bf R}) \rangle = \int d {\bf x}\,  {V}({\bf x}; {\bf R}) \rho({\bf x}| {\bf R})$. This term is minimized by a density, and the corresponding wavefunction, that is localized around the minimum of ${V}({\bf x}; {\bf R})$ ($\rho({\bf x}| {\bf R}) = \delta({\bf x} - {\bar {\bf x}}({\bf R}))$). However, to a very localized wavefunction corresponds a very high ``kinetic energy'', as can be seen by noticing that the ``uncertainty principle'' imposes that $-\langle \psi({\bf x}; {\bf R}) | a \nabla^2 \psi({\bf x}; {\bf R}) \rangle  \geq a / 4  \langle \Delta x^2\rangle$, where $\langle \Delta x^2 \rangle= \langle x^2 \rangle - \langle x \rangle^2$.\cite{Note4}
 Therefore, the wavefunction minimizing the Ritz functional above is the trade off between the need to be localized around the minimum (minima) of the potential $V({\bf x}; {\bf R})$ and the need to be not too localized as, otherwise, the kinetic energy would be too large.  The parameter controlling the degree of localization of the wavefunction is $a$. If $a$ is small, $V({\bf x}; {\bf R})$ is the dominant term and the ground state density is very localized around the absolute minimum of the potential. If, on the contrary, $a$ is large the kinetic energy becomes the dominant term and the ground state density must be highly spread out. 
 
Typically, the kinetic energy of the ground state has to be not too large. This fact has a crucial consequence on the ability of the quantum density representation to distinguish between vacancy and crystal minima also in those cases in which the values of the potential in the two regions are not well separated. To show this, we start by noticing that crystal minima are narrower than vacancy minima. Therefore, the wavefunction corresponding to a $\rho({\bf x}| {\bf R})$ that presents maxima at crystal minima must grow from zero (at the atomic positions) to a large value and go back to zero on a small length scale, corresponding to the inter-atomic distance. A wavefunction of this type will have a large gradient in that region and, therefore, a large kinetic energy, and will not correspond to the ground state of the above Hamiltonian. Therefore, the ground state density will be peaked at the (wider) vacancy minima, also when the  value of $V({\bf x}; {\bf R})$  in the two different types of minima is not very different.
 Indeed, in Sec.~\ref{sec:ResultsAndDiscussion} we will show that $\rho({\bf x}| {\bf R})$ is localized at the vacancy even in the case in which the values of the potential at crystalline and vacancy minima are the same. 
 
The kinetic energy term is also responsible for the smoothness of the ground state density. In fact, using the same argument used for its localization, to a rough wavefunction would correspond a high kinetic energy. Thus, the ground state wavefunction must be smooth and, therefore, the ground state density will be smooth too. 

Summarizing, for a suitable choice of $a$, the quantum density is localized around the vacancy minima of the potential $V({\bf x}; {\bf R})$, even when the value of the potential at the vacancy and crystal minima is similar,  and smooth. 

{The discussion above, and the computer experiments reported in Sec.~\ref{sec:ResultsAndDiscussion-AnalysiOfTheDensity}, indicate that $\rho({\bf x}| {\bf R})$ is} 
able to localize the vacancy and to characterize its migration path. Therefore, we could think of using $\rho({\bf x}| {\bf R})$, or better its value on a discretization of the $x$-space ($\{\rho({\bf x}_i| {\bf R})\}_{i=1,M}$), as a vectorial (field-like) CV in rare event simulations. However, this approach presents problems. First of all, the dimensionality of this vectorial CV, which corresponds to the number of grid points in the discretization of the $x$-space, would be quite large, typically 
much larger than the number of atoms in the simulation sample. This is because we need to characterize accurately the density between atoms. For example, in the simple local migration event shown in Fig.~\ref{fig:FeedbackPotentialAndConfigurations} we need to represent accurately the $\rho({\bf x}| {\bf R})$ in the region between the moving atom and its nearest neighbors, where a new minimum of the potential $V({\bf x}; {\bf R})$ is forming. This requires to have a mesh with several grid points per lattice site, from which we can conclude that the typical dimensionality of the discretized field-like CV proposed above is larger than the number of atoms. It is also worth mentioning that, thanks to the localized nature of the $\rho({\bf x}| {\bf R})$, only a small subset of the elements of this vectorial CV is non negligible at each atomic configuration. In other words, the amount of information provided by $\{\rho({\bf x}_i; {\bf R})\}_{i=1,M}$  can be redundant if we just want to identify the ``position'' of the vacancy and follow its dynamics. 

The use of $\{\rho({\bf x}_i; {\bf R})\}_{i=1,M}$ poses also another, severe, problem: the functional set of the ground state densities of the Hamiltonian ${\mathcal H}({\bf x}; {\bf R})$ is, in general, unknown. This problem, referred to as ``$V$-representability of the ground state densities'' in density functional theory,\cite{GrossDreizler} adds an additional difficulty in using this CV in both guided (umbrella sampling,\cite{UmbrellaSampling} blue moon,\cite{BlueMoon} etc.) and unguided (TAMD/TAMC,\cite{TAMD,TAMC} Metadynamics,\cite{Metadynamics, Metadynamics2} etc.) rare event simulations. 
In the first case, because we do not know how to set {$V$-representable} restraint/constraint values of the CV. {In the second case, since the random value of the CV generated along the biased dynamics might be non $V$-representable, the} simulation will spend a non negligible amount of time sampling values of $\{\rho({\bf x}_i; {\bf R})\}_{i=1,M}$ that are not in the set of the possible solutions, thus reducing the efficiency of the approach. 
A possible, still informative, simplification of the representation consists in using as CVs few low order  moments $\kappa^n$ of the density $\rho({\textbf x}| {\textbf R})$, where $n$ is the order of the moment. While the full representation of the density requires the infinite set of its moments $\{\kappa^n\}_{n=1,\infty}$,\cite{moments} its low order terms, $\kappa^1({\bf R}) = {\bar {\bf x}}({\bf R}) \equiv \int d{\textbf x}\,\, \rho({\bf x}| {\bf R}) {\bf x}$ and $\kappa^2_{\alpha,\beta}({\bf R}) = {\hat c}^2_{\alpha,\beta}({\bf R}) = \int d{\textbf x}\,\, \rho({\bf x}| {\bf R})  (x^\alpha - {\bar x}^\alpha)(x^\beta - {\bar x}^\beta)$, with $\alpha, \beta = 1, \dots, d$ ($d$ dimensionality of the space), seem adequate to catch the features of $\rho({\textbf x}| {\textbf R})$ characterizing the state of the vacancy. The first moment ${\bar {\bf x}}({\bf R})$ tells us where the vacancy is located, whether on a lattice site or in between sites, the latter case indicating that a migration event is taking place. 
Taken alone, ${\bar {\bf x}}({\bf R})$ is not sufficient to characterize the vacancy migration process.  For example, the ${\bar {\bf x}}({\bf R})$ cannot distinguish between a local vacancy migration event, involving only one atom, from a collective one, with several atoms moving at the same time from their initial site to the next one, resulting in a long range migration of the vacancy. 
The second moment can help distinguishing between these two kinds of processes because ${\hat c}^2({\bf R})$ measures the {width of the vacancy, which depends on the number and positions of the sites involved in the process.} In particular, the trace of this matrix, $Tr[{\hat c}^2({\bf R}) ]$, is small if the migration is local and large if the migration is long range. 

In conclusion, the collective variables we propose to use to study the vacancy migration process are the first moment and the trace of the second moment matrix of the density $\rho({\bf x}| {\bf R})$. Should this set of collective variables result insufficient, {e.g. should the anisotropy or the asymmetry of the $\rho({\bf x}| {\bf R})$ be a feature characterizing the vacancy migration in the system at hand,} the set could be improved by replacing the trace of the second moment with the entire matrix (or its eigenvalues), or by adding moments of higher order{, for example the skewness}. The full {approach, i.e. the use of $\rho({\bf x}| {\bf R})$ as CV,} would still be desirable, but this requires further, nontrivial, thinking to solve the problems mentioned before.

\section{Results and discussion}
\label{sec:ResultsAndDiscussion}
This section is divided in two subsections. In Sec.~\ref{sec:ResultsAndDiscussion-AnalysiOfTheDensity} we show that the quantum density representation is able to describe the vacancy migration process at zero and finite temperature, both when the potential in the crystal and vacancy minima has well separated values (the WCA potential will be used as a representative example) and when these values are superimposed (hard disks). In Sec.~\ref{sec:ResultsAndDiscussion-TAMC} we study the vacancy diffusion in a 2D crystal using the reduced set of collective variables introduced at the end of Sec. \ref{sec:QuantumProbeParticle}, and exploit these data to identify possible diffusion paths.

\subsection{Use of the quantum probe quasi-particle representation for the identification of a vacancy in a 2D crystal.}
\label{sec:ResultsAndDiscussion-AnalysiOfTheDensity}
The WCA pair potential is a purely repulsive potential of the form:
\begin{equation}
v(r) = \left \{
\begin{array}{l}
4 \epsilon \left [ {(\sigma / r)}^{12}  - {(\sigma / r)}^{6}\right ] + \epsilon, \,\,\forall\, r < 2^{1/6} \sigma \\
0,  \,\,\forall\, r \geq 2^{1/6} \sigma  
\end{array}
\right .
\end{equation}

\noindent For uniformity of notation, we will denote by $\sigma$ also the size of the hard disks.
The simulations are performed on a sample of 24 atoms and one vacancy in a {periodic} trigonal 2D box, corresponding to one defected $5 \times 5$ layer of the (111) surface of a face centered cubic lattice (see Fig.~\ref{fig:FeedbackPotentialAndConfigurations}). For the WCA system, the lattice constant was fixed to $1.075 \sigma$, to be compared with $2^{1/6} \sigma  \sim 1.222~\sigma$, the value at which the WCA potential becomes zero.  For the hard disk system, we set the lattice constant to $2.5~\sigma$.

Before moving to the description of the calculation of $\rho({\bf x}| {\bf R})$ we must explain how to set the value of $a$. In Sec.~\ref{sec:QuantumProbeParticle} we explained that $a$ must be neither too small, as otherwise the $\rho({\bf x}| {\bf R})$ will be too rough and could not distinguish between vacancy and crystal minima, nor too large, as otherwise the density will tend to be uniformly distributed over the entire $x$-space. 
In practice, we set its value such that the width of $\rho({\bf x}| {\bf R})$ ($\Delta x = \sqrt{\langle ({\bf x} - \langle{\bf x}\rangle )^2\rangle }$) in a crystal with a vacancy at the equilibrium configuration is of the order of the typical interactomic distance in the crystal, e. g. the size of the unit cell for Bravais lattices, or the Van der Waals radius of the missing atom for more complex crystals. In the present work we set $a = 1500$. Below, we investigate the effect of $a$ on the $\rho({\bf x}| {\bf R})$ and show that this value is adequate for the WCA system.

The $\psi({\textbf x}; {\textbf R})$ is computed by expanding the wavefunction on a plane wave basis set  of elements $\chi_i({\textbf x}) = 1/\sqrt{V} \exp[i{\bf G}_i{\bf x}]$ {(see Ref.~\onlinecite{MarxHutter} for technical details of planewave calculations)}. 
The wavevectors ${\bf G}_i$ satisfy the usual condition that the corresponding $\chi_i({\textbf x})$ is periodic over the simulations box, which amounts to set the $\{{\bf G_i}\}_i$ to the points of the reciprocal lattice of the simulation box (${\bf G}_i = 2 \pi {\hat H}^{-1} {\bf u}_i$, where ${\hat H} = [{\bf a}\, \, {\bf b}]$, with ${\bf a}$ and ${\bf b}$ representing the edges of the simulation box, and $u_ i = \pm 1, \pm 2, \dots$). The expansion is limited to the planewaves satisfying the condition $a |{\bf G}_i|^2/2 \leq E_{cut}$. $E_{cut}$ is fixed such that the ground state eigenvalue of the Hamiltonian matrix at several atomic configurations is well converged, i.e. the variation with $E_{cut}$ is less than $10^{-3}~\Delta\epsilon$, with $\Delta\epsilon = \epsilon_1 - \epsilon_0$ and $\epsilon_0$ and $\epsilon_1$  the eigenvalues of the ground and first exited state of the Hamiltonian, respectively. In the present case, $E_{cut}$ was set to $5 \times 10^6$, corresponding to a $100 \times 78$ points Fourier mesh. 

The ground state of ${\mathcal H}({\bf x}; {\bf R})$ {in the $\Gamma$-point approximation} is obtained using the Lanczos iterative method.\cite{Lanczos} In particular, we used the implicitly restarted version of the method,\cite{Arnoldi,arpack} which is more efficient than the original one.

The fact that the WCA and hard disk potentials diverge at ${\textbf x} = {\textbf R}_i$ and $|{\textbf x} - {\textbf R}_i| \leq \sigma$, respectively, poses a problem for the calculation of the matrix elements of the Hamiltonian in the planewave basis set. This problem was solved by replacing the original pair potentials with non diverging approximations. In the WCA case, the original pair potential is replaced with the following approximation:

\begin{equation}
\label{eq:potApproximated}
{\tilde v}(r) = \left \{
\begin{array}{c}
 {v}(r_{cut}), \forall r \leq r_{cut} \\
 {v}(r), \forall r > r_{cut}
 \end{array} \right .
\end{equation}

\noindent We tested the dependency of our results on $r_{cut}$ and verified that for a small enough value of this parameter, such that $V({\textbf x}; {\textbf R}) \equiv {\tilde V}({\textbf x}; {\textbf R}) $ around the vacancy minima, the results are independent from the chosen value of $r_{cut}$ (in the present work $r_{cut}$ was set to $0.35$). If a more accurate approximation were needed, it would be possible to develop a pseudopotential-like approach.\cite{Martin} For the hard disk case, we solved the problem by setting the value of the potential to a large but finite value for $r \leq \sigma$. In the calculations discussed in the following, we set this value to $10$.  Also in this case, we tested that above a given threshold our results are independent on the specific value chosen. 

{In our calculations we also need to ensure that the vacancy wavefunction does not interacts with its periodic images. To test this, we computed the dependency of $\rho({\bf x}| {\bf R})$ on the sample size for selected configurations  along the ideal local migration path (Fig.~\ref{fig:FeedbackPotentialAndConfigurations}). We measured this dependency by computing two quantities: i) $\max_{\bf x} \Delta \rho^{i \times i}({\bf x}| {\bf R}) / \max_{\bf x} \rho^{5 \times 5}({\bf x}| {\bf R})$, where $\Delta \rho^{i \times i}({\bf x}| {\bf R}) = | \rho^{i \times i}({\bf x}| {\bf R}) - \rho^{5 \times 5}({\bf x}| {\bf R})|$ and  $\rho^{i \times i}({\bf x}| {\bf R})$ is the vacancy density in a $i \times i$  system, and ii) $\int\,\, d{\bf x}\,\, \Delta \rho^{i \times i}({\bf x}| {\bf R})$. For $i = 10 -30$ (with increments of $5$) we found that $\max_x \Delta \rho^{i \times i}({\bf x}| {\bf R}) / \max_x \rho^{5 \times 5}({\bf x}| {\bf R}) \leq 5 \times 10^{-4}$ and $\int\,\, d{\bf x}\,\, \Delta \rho^{i \times i}({\bf x}| {\bf R}) \leq 10^{-5}$. The very small dependence of $\rho({\bf x}| {\bf R})$ on the size of the simulation box is due to its very localized nature.

We start the presentation of our results by showing that the quantum density description is able to localize the vacancy in the correct region of the space  for the well ordered atomic configurations reported in Fig.~\ref{fig:FeedbackPotentialAndConfigurations} (corresponding to $T = 0$), both in the case of the WCA and hard disk pair potentials. 
In 
Fig.~\ref{fig:handPreparedConfigurations-Density} we plot the $\rho({\bf x}| {\bf R})$ associated to the configurations of 
Fig.~\ref{fig:FeedbackPotentialAndConfigurations}. 
{Initially, when the atoms are at their equilibrium position (top panel), the density is localized at the vacancy site. Then, while the migrating atom proceeds along its linear path, the density follows the vacancy minimum of the $V({\bf x}; {\bf R})$ (the two central panels), finally splitting in a symmetric bimodal $\rho({\bf x}| {\bf R})$ when the migrating atom is in the mid-way configuration (bottom panel).}
This figure clearly shows that the quantum density representation is able to describe the vacancy all along the (ideal) local migration path in a system in which there is a sizeable difference in the value of $V({\bf x}; {\bf R})$ between vacancy and crystal minima. However, as explained in Sec.~\ref{sec:QuantumProbeParticle}, the quantum density representation is able to characterize the vacancy also when there is no such a separation. This is shown in Fig.~\ref{fig:handPreparedConfigurationsHD-Density}, where we report data analogous to those of Fig.~\ref{fig:handPreparedConfigurations-Density} for a system of hard disks (only data for the equilibrium and mid-way configurations are shown). Also for the hard disk system the quantum density representation is able to properly describe the vacancy and its dynamics along the ideal local migration path. 

{To be able to use the quantum probe quasi-particle description we need to show that it is possible to find a values of the parameter $a$  such that $\rho({\bf x}| {\bf R})$, and its low order moments considered before, are able to identify the vacancy position and follow its migration. This is verified using ${\bar {\bf x}}({\bf R})$ and $Tr[{\hat c}^2({\bf R}) ]$ computed at selected configurations along the local migration path (Fig.~\ref{fig:FeedbackPotentialAndConfigurations}). The criterium we adopt is that the description is valid if the variation of the values of these observables computed at significantly different configurations along the migration paths (e.g. the initial and the mid-way configuration of Fig.~\ref{fig:FeedbackPotentialAndConfigurations}) are larger than thermal fluctuations. For the WCA system at $T = 0.5$, for example, the root mean square thermal fluctuation of the first and second moment, $\Delta {\bar {\bf x}}({\bf R})$ and $\Delta Tr[{\hat c}^2({\bf R}) ]$, are 0.1 and 0.05, respectively (see left panels of Fig.~\ref{fig:collVarTAMCvsMC}). In Fig.~\ref{fig:firstSecondMomentsvsA} we report ${\bar {\bf x}}({\bf R})$ and $Tr[{\hat c}^2({\bf R}) ]$ {\itshape vs} $a$. 
A curve in the top, central and bottom panel of this figure represents ${\bar {\bf x}}_1({\bf R})$, ${\bar {\bf x}}_2({\bf R})$ and $Tr[{\hat c}^2({\bf R}) ]$, respectively, at a given atomic configuration {\emph vs} $a$. Curves with same color and symbol in different panels refer to the same configuration. We identify three zones, denoted $1$, $2$ and $3$ in the figure. In zone $1$ $\rho({\bf x}| {\bf R})$ is essentially independent on the value of  $a$, and ${\bar {\bf x}}({\bf R})$ and $Tr[{\hat c}^2({\bf R}) ]$ are almost constant over the entire zone. This happens when the $\rho({\bf x}| {\bf R})$ is more localized than the interatomic distance ($1.075~\sigma$ in our case) in the vacancy minima. In this zone, the difference of the values of ${\bar {\bf x}}({\bf R})$ and $Tr[{\hat c}^2({\bf R}) ]$ between the initial and mid-way configuration are larger than their thermal fluctuation at $T = 0.5$. Moreover, one of the two components of the first moment, ${\bar x}_2({\bf R})$, is well separated also for configurations closer to each other, e.g. at $1/5$ and $2/5$ of the arclength distance $l$ along the local migration path, which is enough to distinguish between these two different states.  In zone $2$, $\rho({\bf {\bf x}}| {\bf R})$ is no longer independent on $a$ and, for a given configuration,  ${\bar {\bf x}}({\bf R})$ and $Tr[{\hat c}^2({\bf R}) ]$  change continuously with $a$. However, also in this case, for a given value of $a$, the values of ${\bar {\bf x}}({\bf R})$ and $Tr[{\hat c}^2({\bf R}) ]$ at significantly different configurations along the local migration path are well separated. Finally, in the zone $3$, $\rho({\bf x}| {\bf R})$ is no longer localized in the vacancy region and, as a consequence, ${\bar {\bf x}}({\bf R})$ and $Tr[{\hat c}^2({\bf R}) ]$ take  values essentially independent from the atomic configuration. Obviously, values of $a$ belonging to this range are unsuitable. Summarizing, the quantum probe particle description is suitable to represent the vacancy and track its dynamics over a wide range of values of $a$ (i.e. the choice of the value of $a$ is not critical), and the criterion we mentioned before, that $a$ must be such that $Tr[{\hat c}^2({\bf R}) ]^{1/2}$ is close to the interatomic distance, is adequate.}

 \begin{figure}
 \begin{center}
 \includegraphics[width=0.5\textwidth]{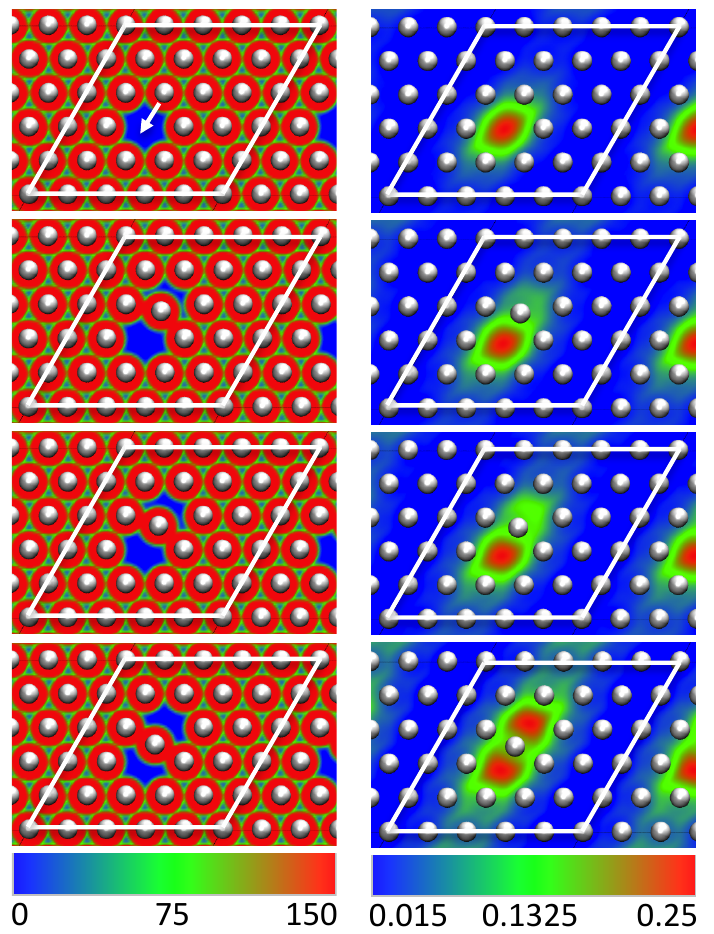}
 \end{center}
 \caption{$V({\bf x}; {\bf R})$ (left) and $\rho({\bf x}| {\bf R})$ (right) for four atomic configurations along the ideal local vacancy migration path in a system of WCA particles. 
 \label{fig:handPreparedConfigurations-Density}}
 \end{figure}

 \begin{figure}
 \begin{center}
 \includegraphics[width=0.450\textwidth]{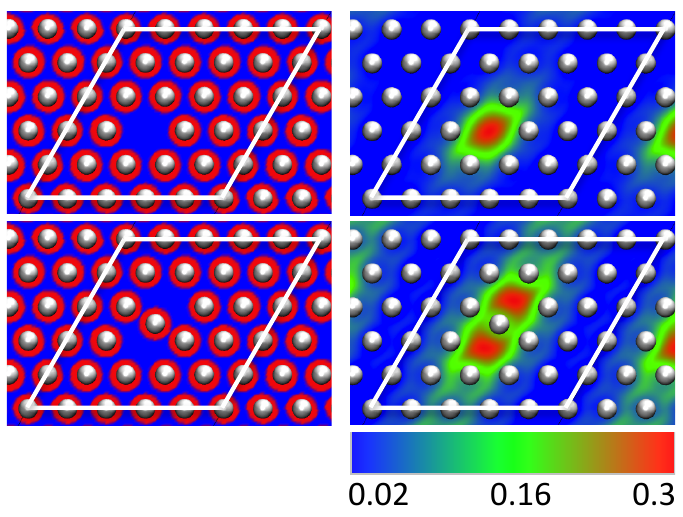}
 \end{center}
 \caption{$V({\bf x}; {\bf R})$ (left) and $\rho({\bf x}| {\bf R})$ (right) for a system of hard disks in the equilibrium and mid-way configurations along the ideal local migration path. The blue color in the left-hand panel denotes  $V({\bf x}; {\bf R}) = 0$; the red  $V({\bf x}; {\bf R}) = \infty$. 
 \label{fig:handPreparedConfigurationsHD-Density}}
 \end{figure} 
 
 \begin{figure}
 \begin{center}
 \includegraphics[width=0.7\textwidth]{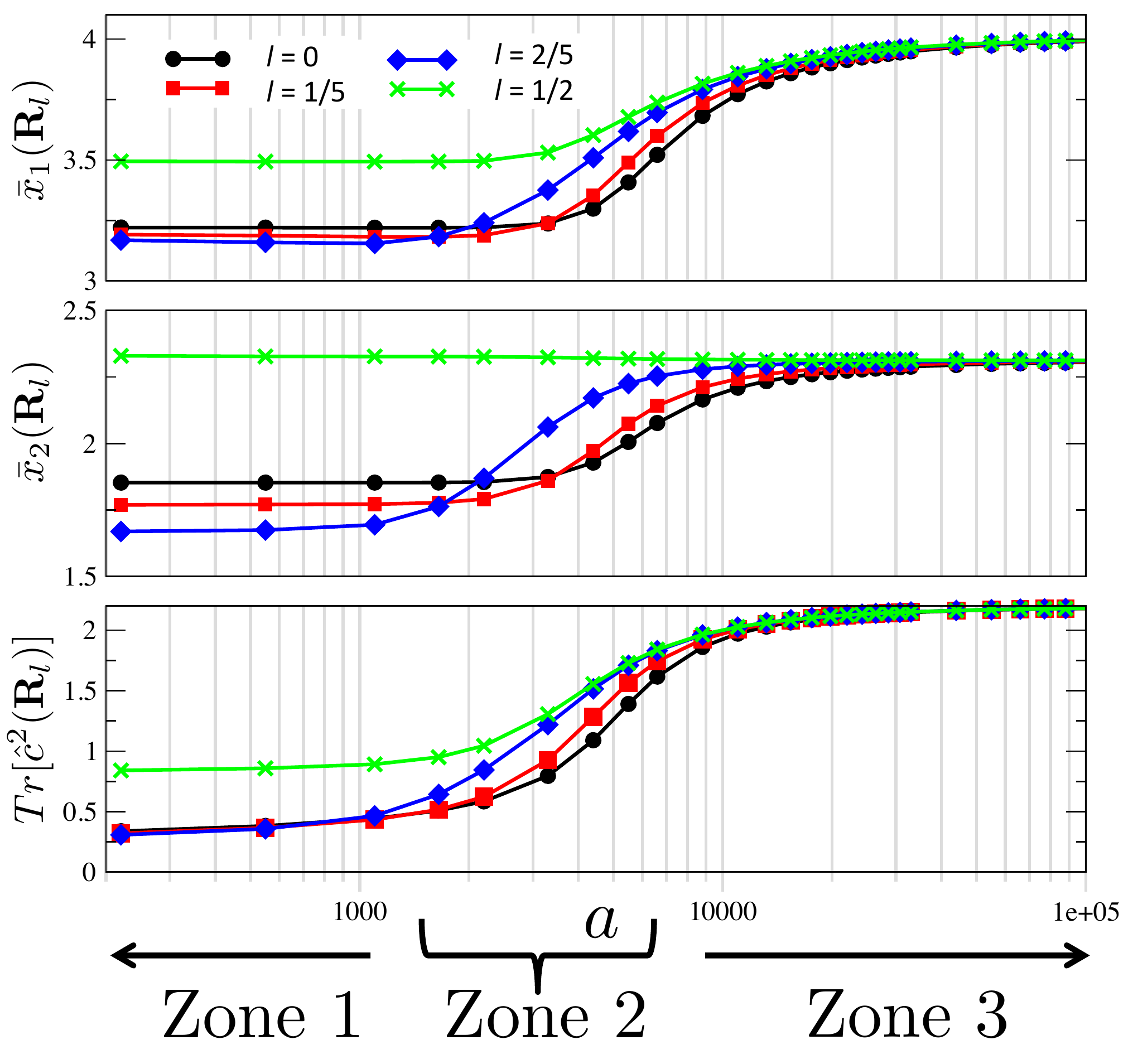}
 \end{center}
 \caption{{${\bar {\bf x}}({\bf R})$ and $Tr[{\hat c}^2({\bf R}) ]$ as a function of $a$ at the configurations shown in Fig.~\ref{fig:FeedbackPotentialAndConfigurations}. $a$ is reported on a logarithmic scale. Vertical lines are inserted to help the comparison of values of ${\bar {\bf x}}({\bf R})$ and $Tr[{\hat c}^2({\bf R}) ]$ at a given value of $a$. $l$ denotes the arclength distance along the ideal local migration path of Fig.~\ref{fig:FeedbackPotentialAndConfigurations} starting from the initial condition. Zone $1$, $2$ and $3$ denote the three regions of $a$ in which ${\bar {\bf x}}({\bf R})$ and $Tr[{\hat c}^2({\bf R}) ]$ have a different behavior (see text).}
 \label{fig:firstSecondMomentsvsA}}
 \end{figure}

In order to test whether this model is still able to represent the vacancy and its dynamics when the system is at finite temperature, we run a high temperature ($T = 2.5$) MD simulation with a Langevin thermostat. In this case, we observed one vacancy migration event in a $10000$-step long simulation. On the short timescale of our MD simulation the system remains crystalline, even though its equilibrium state at this $T$ could be the liquid one. 
In the left column of 
Fig.~\ref{fig:nonAcceleratedVacancy-Density} is reported the potential $V({\bf x}; {\bf R})$ along the (local) migration event mentioned above. By comparing the potential shown in 
Fig.~\ref{fig:nonAcceleratedVacancy-Density} and 
Fig.~\ref{fig:FeedbackPotentialAndConfigurations} we see that, at finite temperature, the $V({\bf x}; {\bf R})$ along a local migration event has still a shape similar to that at $T = 0$. However, few narrow minima, of magnitude similar to those of the vacancy ones, are also present. These minima are due to large displacements of atoms not involved in the vacancy migration process out of their equilibrium position. Despite the more complex shape of the potential in the configurations visited at finite (and high) temperature, the $\rho({\bf x}| {\bf R})$ is still able to correctly identify the position of the vacancy, both when the vacancy is located at a lattice site and when it is moving from one site to another (see the right-hand panel of 
Fig.~\ref{fig:nonAcceleratedVacancy-Density}). Indeed, while some density leaks to the crystalline minima, the value there is negligible compared to the one around the vacancy ones.

 \begin{figure}
 \begin{center}
 \includegraphics[width=0.45\textwidth]{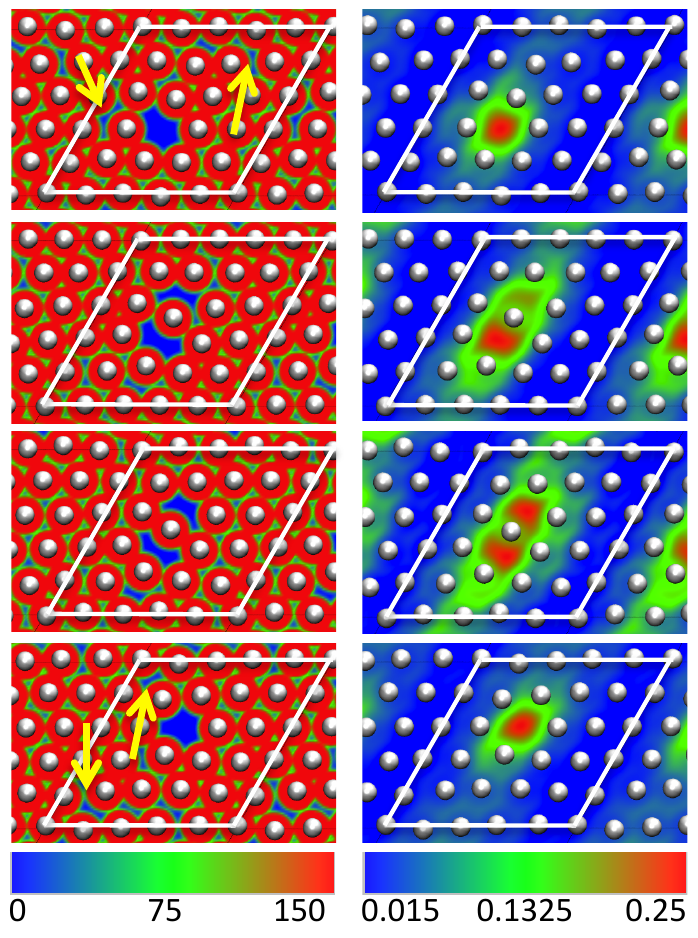}
 \end{center}
 \caption{$V({\bf x}; {\bf R})$ (left) and $\rho({\bf x}| {\bf R})$ (right) at few selected atomic configurations along the migration event observed during an unbiased simulation at $T = 2.5$. The arrows in the left column point to representative narrow minima of the potential $V({\bf x}; {\bf R})$ not associated to the vacancy (see text).}\label{fig:nonAcceleratedVacancy-Density}
 \end{figure}

We also analyzed the high temperature MD trajectory to test the ability of ${\bar {\bf x}}({\bf R})$ and $Tr [{\hat c}^2({\bf R})]$ to monitor the vacancy migration processes.  if our CVs are good,  their oscillations when our system is in a metastable state should be much smaller than their variation when a vacancy migration takes place. This condition is indeed verified, as can be seen by comparing the typical oscillation of the CVs against their (simultaneous) change at the timestep~$\sim 4700$ in the ${\bar {\bf x}}({\bf R})$ and $Tr [{\hat c}^2({\bf R})]$ timelines along the $T = 2.5$ simulation (Fig.~\ref{fig:nonAcceleratedVacancy-FirstSecondMomentum}), corresponding to the migration event.
By visualizing the trajectory, we verified that this level is of local kind (this is, indeed, the event shown in Fig.~\ref{fig:nonAcceleratedVacancy-Density}). In Fig.~\ref{fig:nonAcceleratedVacancy-FirstSecondMomentum} we also notice several peaks in the $Tr [{\hat c}^2({\bf R})]$ curve to which do not correspond any significant change in the ${\bar x}({\bf R})$. We analyzed the origin of this behavior by looking at the $\rho({\bf x}| {\bf R})$ during one of these events, namely the one at the timestep $\sim 5600$ (indicated by an arrow in Fig.~\ref{fig:nonAcceleratedVacancy-FirstSecondMomentum}). A series of snapshots taken along this event are shown in Fig.~\ref{fig:nonAcceleratedVacancyMissedHopping-Density}. The sudden change in the second moment is due to an unsuccessful migration event. This produces a broadening of the  $\rho({\bf x}| {\bf R})$, which partly populates also the minimum of the $V({\textbf x}; {\textbf R})$ that is forming on the crystal site initially occupied by the migrating atom. However, the process does not end with a migration and the final  $\rho({\bf x}| {\bf R})$ is still localized on the initial empty crystal site. The fact that the first moment does not change (significantly) during this event is due to the combination of two factors.
 On the one hand, during the attempted migration event there is an increase of the density on the site of the moving atom. This should move the center of the density $\rho({\bf x}| {\bf R})$, i.e. ${\bar {\bf x}}({\bf R})$, toward this site. However, at the same time, the maximum of the density on the original vacancy site moves in the opposite direction, following the position of the original vacancy minimum of the potential $V({\textbf x}; {\textbf R})$, which is ``pushed'' in this direction by the moving atom. This second effect would move the ${\bar {\bf x}}({\bf R})$ in the opposite direction and the two effects, essentially, compensate. This can be seen in Fig.~\ref{fig:nonAcceleratedVacancyMissedHopping-Density}, where, together with the snapshots of the potential  $V({\textbf x}; {\textbf R})$ (left) and the density $\rho({\bf x}| {\bf R})$ (right) along the attempted migration event, we report the first moment ${\bar {\bf x}}({\bf R})$ (denoted by the white cross), see also the discussion in the caption.

In conclusion, the high temperature MD test indicates that ${\bar {\bf x}}({\bf R})$ and $Tr [{\hat c}^2({\bf R})]$ are both needed to monitor the vacancy migration process. At the same time, this test makes us confident that these three CVs are sufficient to study this process using rare event techniques. 

 \begin{figure}
 \begin{center}
 \includegraphics[width=0.5\textwidth]{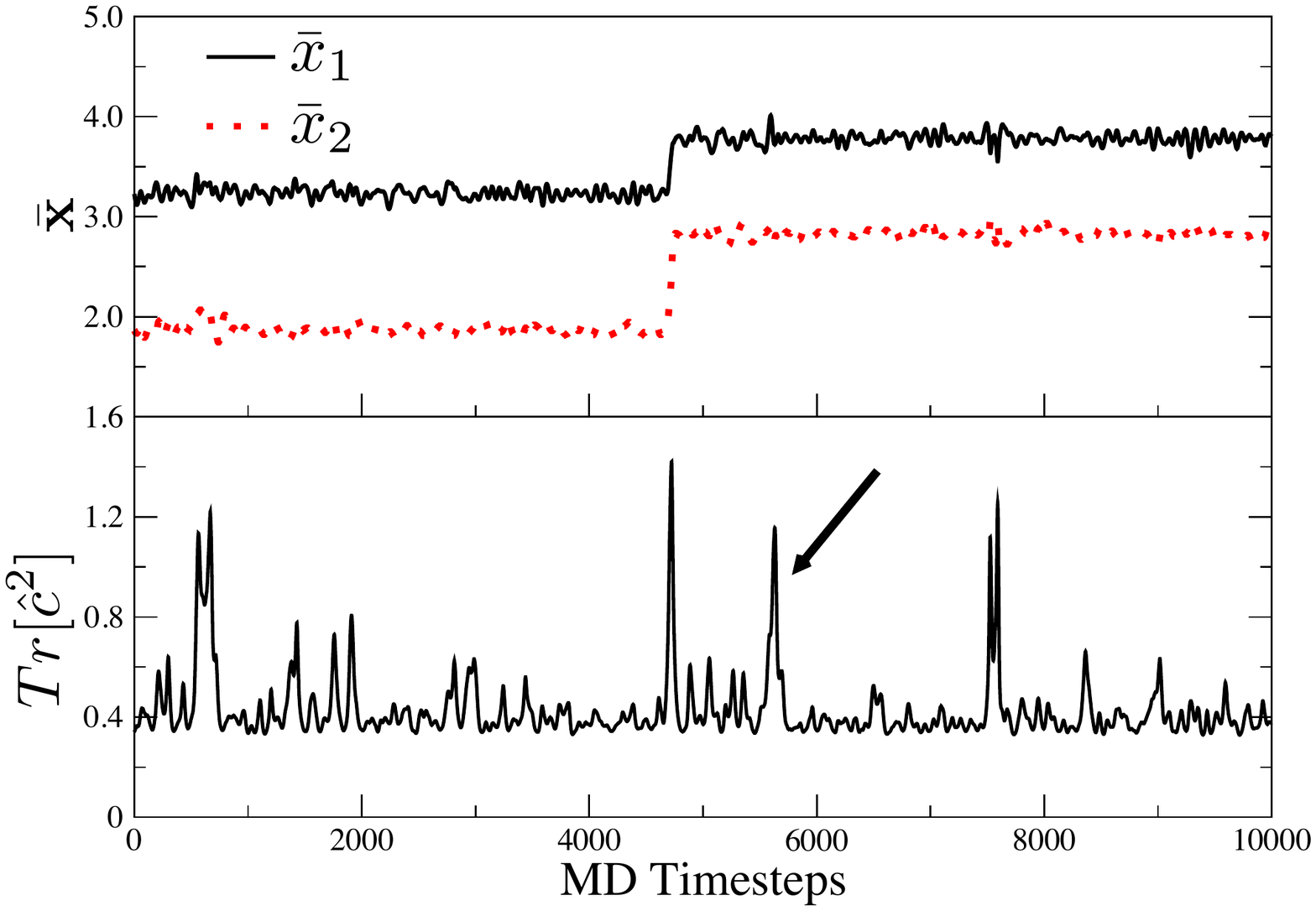}
 \end{center}
 \caption{The two components, ${\bar x}_1$ and ${\bar x}_2$, of the first moment (top) and the trace of the second moment (bottom) of the density $\rho({\bf x}| {\bf R})$ as measured along a high temperature ($T = 2.5$) simulation.}\label{fig:nonAcceleratedVacancy-FirstSecondMomentum}
 \end{figure}
 
\begin{figure}
\begin{center}
\includegraphics[height=0.5\textheight]{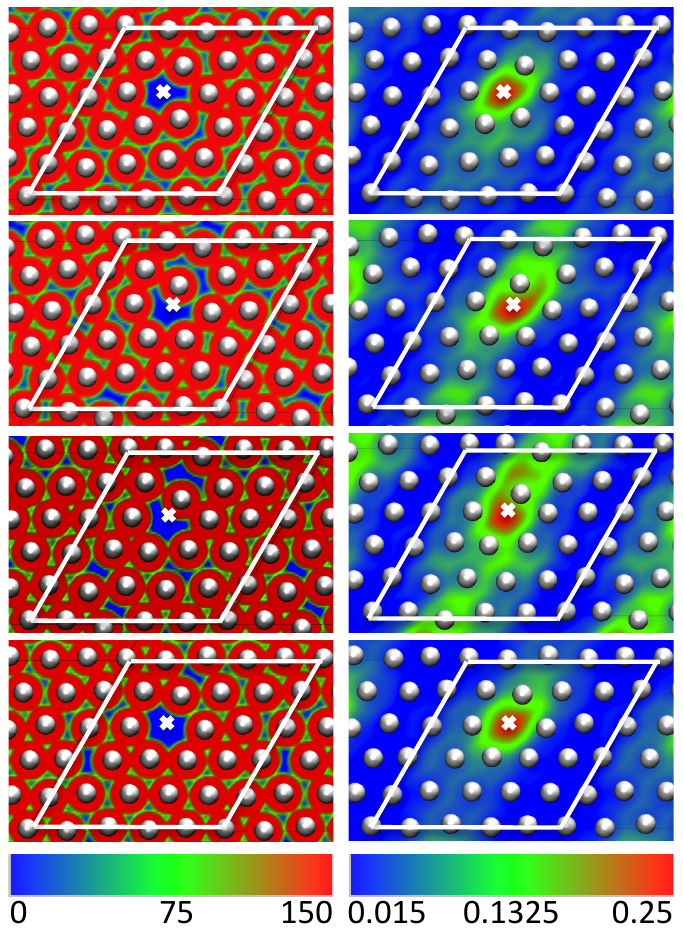}
\end{center}
\caption{Potential $V({\textbf x}; {\textbf R})$ (left) and density $\rho({\bf x}| {\bf R})$ (right) computed on configurations corresponding to few snapshots along the avoided migration event at the timestep $\sim 5600$ of the high temperature MD simulation. The white cross indicates the position of ${\bar {\bf x}}({\bf R})$. When the vacancy is located at a crystal site (top and bottom panels), ${\bar {\bf x}}({\bf R})$ is approximately in correspondence of the maximum of the $\rho({\bf x}| {\bf R})$ and the vacancy minimum of the potential  $V({\textbf x}; {\textbf R})$. During the attempted migration event (central panels), the main peak of the density is shifted toward the bottom-left with respect to ${\bar {\bf x}}({\bf R})$, while a long tail is formed at its top-right. }\label{fig:nonAcceleratedVacancyMissedHopping-Density}
\end{figure}

\subsection{Vacancy diffusion path in a 2D WCA crystal by Temperature Accelerated Monte Carlo.}
\label{sec:ResultsAndDiscussion-TAMC}

In this section we report the results of rare event simulations aimed at identifying vacancy migration paths in a 2D WCA crystal at finite temperature using ${\bar {\bf x}}({\bf R})$ and $Tr [{\hat c}^2({\bf R})]$ as CVs. These calculation will be performed using TAMC.\cite{TAMC,Note5} 
{In TAMC the dynamical system consists of the original atomistic variables plus a set of extra variables ${\textbf z} = \{z_i\}_{i=1,m}$,} associated to a set of collective variables $\{\theta_i({\textbf R})\}_{i=m}$, ${\bar {\bf x}}({\bf R})$ and $Tr [{\hat c}^2({\bf R})]$ in the present case. The two sets are coupled via the potential energy term $U_k({\bf z}, {\bf R}) = \sum_{i=1,m} k_i/2 (\theta_i({\textbf R}) - z_i)^2$. The atoms and the $\textbf z$ are ``evolved'' together, the atoms according to a standard Metropolis Monte Carlo governed by the physical potential energy plus the term $U_k({\bf z}, {\bf R})$, and the $\bf z$ according to a constant temperature dynamics (Langevin dynamics in the present case) governed by the potential $U_k({\bf z}, {\bf R})$. The inertia of the $\bf z$, a tunable parameter in this method, can be set such that the evolution of these variables is adiabatically separated from the Monte Carlo on the atoms, so that the latter samples the conditional probability density function $m({\bf R}| {\bf z})$. It can be shown (see Ref.~\onlinecite{TAMC}) that under these conditions the variables $\bf z$ sample the probability density function $P_{\theta}({\bf z_i}) = \exp[-\beta^* F_k({\bf z})]$, where $\beta^* = 1 / k_B T^*$ ($k_B$ Boltzmann constant) and, for $k$ large enough, $F_k({\bf z})$ is the free energy {at the physical temperature $T$} associated to the state $\theta({\textbf R}) = {\bf z}$. $T^*$ is the temperature of the ${\bf z}$ variables, which can be different from the physical temperature. By defining $T^*$ such that the associated thermal energy is higher than the barriers separating free energy minima, we are able to efficiently explore the free energy surface. 

In this work, we want to identify possible vacancy migration paths in a 2D WCA crystal by direct inspection of microscopic configurations explored in our TAMC runs.
The reconstruction of the free energy surface, {the accurate determination of the migration paths,}  and the corresponding rates via the transition state theory with dynamical corrections,\cite{TST1,TST2} will be the objective of  a forthcoming study on a more realistic 3D system. 

We performed a $2.2 \times 10^6$ steps TAMC simulation on the already described 2D WCA crystal with one vacancy. The physical temperature was set to $T = 0.5$ and the temperature of the CVs to $T^* = 8.5$. The simulation was started from a configuration in which the vacancy and all the atoms are at the lattice sites. In 
Fig.~\ref{fig:collVarTAMCvsMC} we compare the values of ${\bar {\bf x}}({\bf R})$ and $Tr [{\hat c}^2({\bf R})]$ as obtained from a MC simulation at $T = 0.5$ with those obtained from the TAMC simulation. We did not observe any migration event in the MC simulation. This is consistent with recent results~\cite{Pinski2010} on vacancy migration in a closely related 2D crystal at $T = 0.1$, for which the energy barrier of the local migration process was estimated to be $\sim 20$. At variance with MC simulations, in the TAMC run we observed several migration events, identified by the simultaneous change of ${\bar {\bf x}}({\bf R})$ and $Tr [{\hat c}^2({\bf R})]$. In the TAMC simulation, we also observe many peaks in the $Tr [{\hat c}^2({\bf R})]$ not associated to any change of ${\bar {\bf x}}({\bf R})$. As explained in the previous section, these peaks are due to aborted migration events, which at $T^* = 8.5$ are quite frequent. By visual inspection of the atomic ``trajectory'' of the TAMC simulation, we identified three different kinds of events described in the following. 
The first one corresponds to a local migration event. The atomic trajectory along one of the eight events of this type observed in our simulations, namely the one at the step $\sim 1.1 \times 10^6$ (see Fig.~\ref{fig:collVarTAMCvsMC}), is shown in 
Fig.~\ref{fig:Diffusion}/A. It can be noticed that while the atom $1$ is migrating toward the vacancy the next neighbor atom (atom $2$ in the figure) is moving in the same direction, giving rise to a concerted motion. This kind of synchronous motion is found also in the other local migration events observed in the TAMC simulation.

\begin{figure}
\begin{center}
\includegraphics[width=0.5\textwidth]{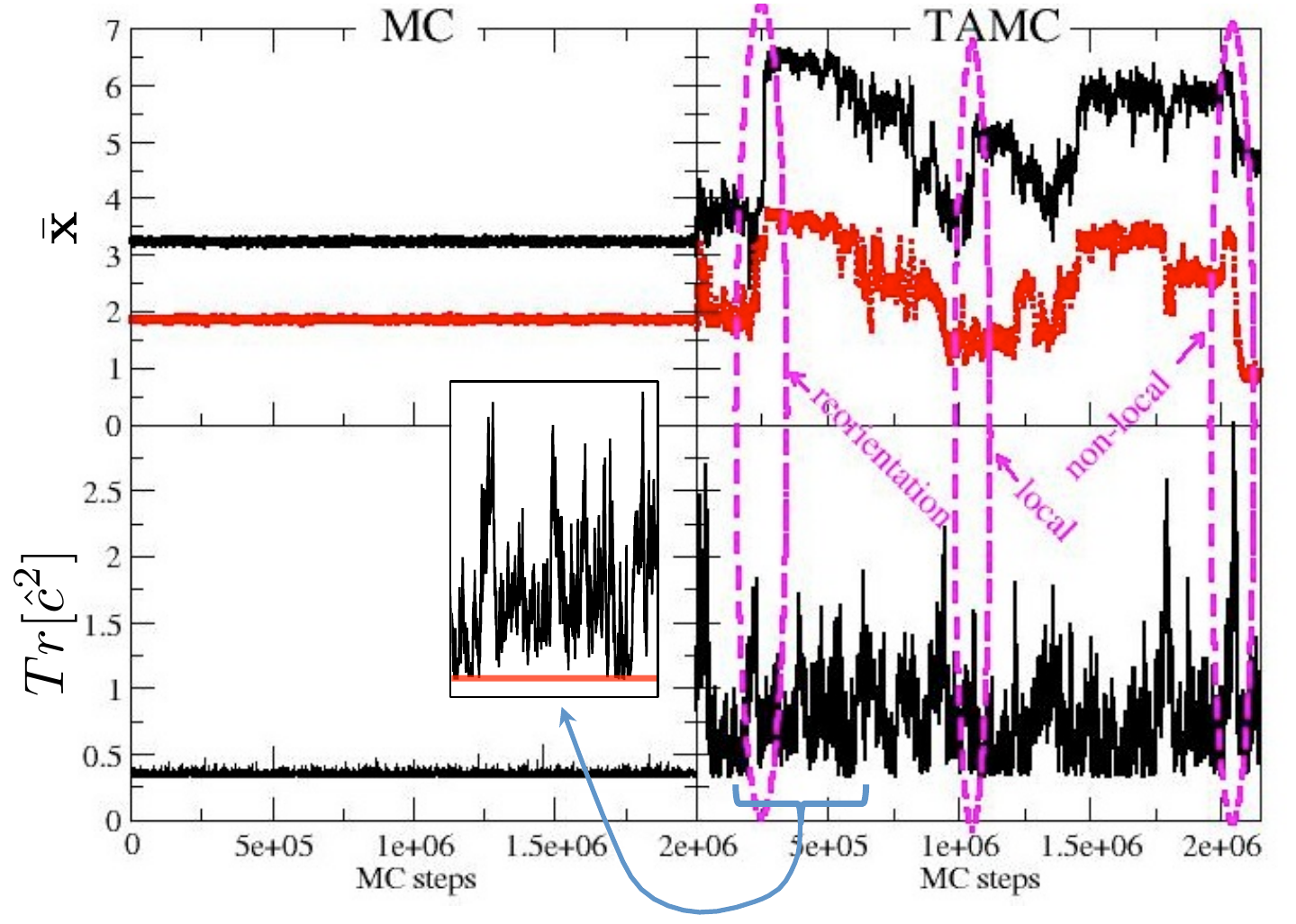}
\end{center}
\caption{Comparison of the first moment and the trace of the second moment of the density $\rho({\bf x}| {\bf R})$ as obtained from (unbiased) MC at $T = 0.5$ (left) and TAMC at the same $T$ and at $T^* = 8.5$ (right). The first component of the $\bar {\bf x}$ ($\bar {x}_1$) is denoted by a continuous - black - line, while the second component ($\bar {x}_2$) by a dotted - red - line. The dashed ellipses indicate events that are discussed in the text. The inset shows a zoom of the $Tr [{\hat c}^2({\bf R})]$  along the TAMC simulation in the interval in which the system was in a reoriented crystalline state (see text). The red line correspond to the bottom of the $Tr [{\hat c}^2({\bf R})]$ curve in the rest of the simulation.}\label{fig:collVarTAMCvsMC}
\end{figure}
 
\begin{figure*}
\begin{center}
\includegraphics[width=0.95\textwidth]{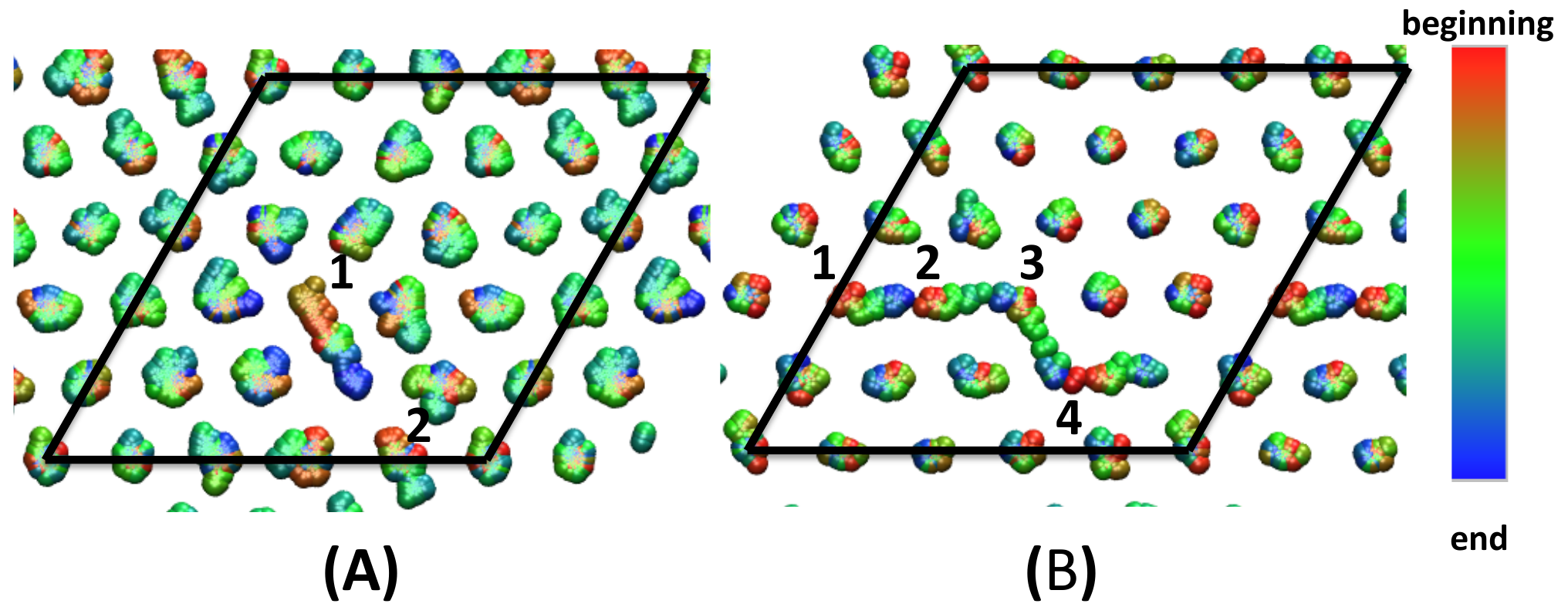}
\end{center}
\caption{Trajectories along a local (A) and non-local (B) migration event. The particles are colored in red at the beginning of the trajectory and, passing by green, become blue at the end. The boundary of the simulation box is also reported.}\label{fig:Diffusion}
\end{figure*}
 
\begin{figure*}
\begin{center}
\includegraphics[width=0.95\textwidth]{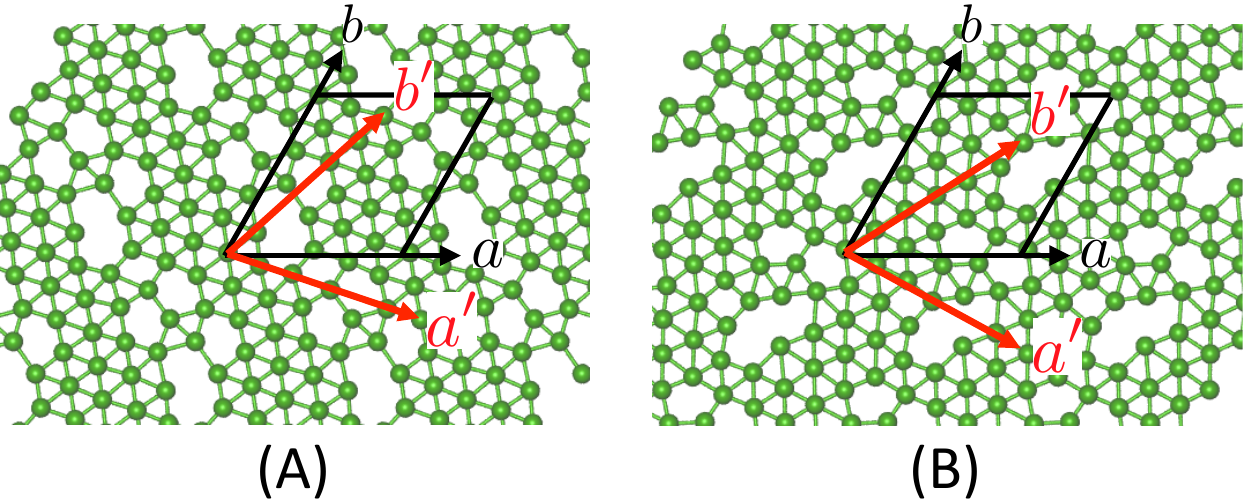}
\end{center}
\caption{Atomic configurations corresponding to the misoriented crystal. The stress due to wrong orientation is accommodated by the formation of a wide vacancy defect. During the interval $\sim 2.3 - 5 \times 10^5$ TAMC steps the misoriented crystal kept the same geometry (trigonal) but its orientation and the position of the vacancy changed several times. The misorientation of the crystal is put in evidence by reporting on the plot the original lattice vectors ${a}$ and ${b}$, and the lattice vectors of the reoriented crystal ${a}'$ and ${b}'$.}\label{fig:changeOrientation}
\end{figure*}

In 
Fig.~\ref{fig:Diffusion}/B we report another kind of event, corresponding to the step $\sim  2 \times 10^6$ of the simulation, that we call non local migration event. In this case, several atoms and lattice sites are involved in the process, namely four atoms and five lattice sites. The trajectory reported in the figure shows that this event does not consist of a series of independent local migration events close in time. Rather, the atoms move all together and the process ends when they have all reached the final lattice position. The result of this event is that the vacancy is transferred at a distance of four lattice sites from its initial position. This is, indeed, a concerted ``multiple jump'' process of the kind identified by Da Fano and Jacucci in high temperature Na and Al samples.\cite{DaFanoJacucci} During the TAMC simulation, we observed only one event of this type, against eight local migration events. This is consistent with the observation of Da Fano and Jacucci that the non local migration process is statistically relevant only at high temperature (i.e. close to the melting).

Finally, the third kind of process observed in our simulations corresponds to a reorientation of the lattice. This process, taking place at the step $\sim 2.3 \times 10^5$ of the TAMC run, starts as a non-local migration event from a properly oriented crystal structure. This initial step is followed by a global change of the atomic positions into a lattice of analogous symmetry (trigonal) but with a different orientation (see Fig.~\ref{fig:changeOrientation}). This misoriented lattice is incommensurable with the simulation box, thus producing a large stress on the system. The system reacts to this stress by forming wider vacancies, such as those shown in   Fig.~\ref{fig:changeOrientation}. The formation of these defects is reflected on the value of the $Tr [{\hat c}^2({\bf R})]$. In fact, in the inset of Fig.~\ref{fig:collVarTAMCvsMC} you can see that in the interval $\sim 2.3 - 5 \times 10^5$ the bottom of the $Tr [{\hat c}^2({\bf R})]$ curve is higher than in the rest of the simulation, when the system is in the ordinary orientation. The process ends with another global change of the atomic position restoring the original orientation of the crystal. This process is most likely an artefact of the large disorder present in the crystal due to the high vacancy concentration in the sample, which is orders of magnitude higher than the typical value in bulk systems. Nevertheless, it is very promising that our CVs are able to accelerate such a process. This could allow to investigate the formation of similar states/processes in low dimensional systems  (e.g. interfaces, nanocrystals) in which the formation of misoriented crystals might be more favorable than in bulk systems.

\section{Conclusions}
\label{sec:Conclusions}
In this paper we have introduced a novel field-like observable able to locate a single vacancy in a crystal and follow its dynamics. This observable is the ground state probability density of a quantum probe quasi-particle for the Hamiltonian associated to the potential energy field generated by the atoms in the sample. To exploit this observable in practice, we derived from it a small set of collective variables that, used in conjunction with rare event techniques, allowed us to study possible vacancy migration paths in a 2D crystal of Week-Chandler-Andersen particles. Our simulations revealed, in addition to the simple nearest neighbor vacancy migration mechanism, a long range migration path consisting of the simultaneous jump of several atoms. Moreover, we observed a crystal reorientation process induced by a multiple jump vacancy migration event. Work is in progress to generalize this description to the many vacancy case and to the possible creation/annihilation of vacancy/interstitial pairs.

\acknowledgements
We would like to thank Sara Bonella for the careful reading of the manuscript, and the many stimulating discussions and suggestions.
SM acknowledges financial support from the European Community under the Marie Curie Intra-European Fellowship for Career Development Grant No. 255406. GC acknowledges financial support from SFI Grant No. 08- IN.1-I1869 and the Istituto Italiano di Tecnologia under the SEED project grant No. 259 SIMBEDD - Advanced Computational Methods for Biophysics, Drug Design and Energy Research.

\appendix

 \begin{center}
    {\bf APPENDICES}
  \end{center}

\section{Inadequacy of simpler representations.} 
\label{App:simplerApproaches}

One might think of using simpler representation of the vacancy. In this appendix we describe two of them and discuss their inadequacy.

The first one consists in representing the vacancy as a classical probe particle. 
In this case, the vacancy is represented by a particle positioned at ${\tilde {\bf x}}({\bf R}) = \arg \min_{\bf x} V({\bf x}; {\bf R})$,
i.e. the mechanical most stable equilibrium position of a probe particle in the field generated by the atoms. 
This works when the vacancy is at a lattice site but it is no longer adequate when a migration process takes place, as we illustrate by the following ideal experiment of a local vacancy migration event. For a 2D WCA crystals or similar systems, when a vacancy is located at a lattice site the corresponding minimum of $V({\bf x}; {\bf R})$ is much deeper than the crystal minima (see top panel of Fig.~\ref{fig:FeedbackPotentialAndConfigurations}) and the classical probe particle description 
is univocal and adequate. When the atom is mid-way along this path (bottom panel of Fig.~\ref{fig:FeedbackPotentialAndConfigurations}/D),  the potential $V({\bf x}; {\bf R})$ presents two minima of equal value in the region of the two lattice sites involved in the process. In this configuration, the definition of ${\tilde {\bf x}}({\bf R})$ becomes ambiguous. However, even before reaching this double-minimum state, the vacancy migration representation given by the classical probe particle description is unsatisfactory. In fact, while the atom moves along the path described above, a second minimum starts forming close to the original lattice of the migrating atom (two central panels of Fig.~\ref{fig:FeedbackPotentialAndConfigurations}). The magnitude of this minimum increases while the atom moves toward the empty site, until it matches the magnitude of the other minimum in the mid-way configuration, but ${\tilde {\bf x}}({\bf R})$ remains unchanged. Thus, the probe particle description is unable to represent the continuous evolution from the initial to the mid-way configuration of the system.
 
A natural improvement over the previous description could be obtained by describing the vacancy in terms of the (classical) probability density $\rho_C({\bf x}| {\bf R})$ to find a probe particle at ${\bf x}$ conditional to the atoms to be at the configuration ${\bf R}$:

\begin{equation} 
\label{eq:ColVarClassDensity}
\rho_C({\bf x}| {\bf R}) = \frac{\exp[-{\bar \beta} V({\bf x}; {\bf R})]}{ \int dx \exp[-{\bar \beta} V({\bf x}; {\bf R})]}
\end{equation} 

\noindent In 
Eq.~\ref{eq:ColVarClassDensity} ${\bar \beta}$ is a tunable inverse ``temperature'' controlling the localization of $\rho_C({\bf x}| {\bf R})$: the higher is ${\bar \beta}$ the more $\rho({\bf x}| {\bf R})$  is localized around the deeper minimum/minima of the potential $V({\bf x}; {\bf R})$.  $\rho_C({\bf x}| {\bf R})$ is able to correctly represents configurations such as the one shown in the bottom panel of Fig.~\ref{fig:FeedbackPotentialAndConfigurations}, where the vacancy is split in two, when the difference between the values of $V({\bf x}; {\bf R})$ at vacancy and crystal minima is sizeable and when the system is at $T = 0$. However, when these two conditions are not met also the classical density representation has problems. To illustrate this, let us consider first the case discussed in Sec.~\ref{sec:ResultsAndDiscussion-AnalysiOfTheDensity} of a system composed of hard disks. In this case the classical density representation would be completely inadequate since, due to the fact that the potential $V({\bf x}; {\bf R})$ is zero everywhere apart within the disks, the density would be uniform outside the disks. 

Let us now consider finite temparature effects. In this case, atoms move out of their lattice sites, $V({\bf x}; {\bf R})$ will have a more complex landscape than in the $T = 0$ case, and $\rho_C({\bf x}| {\bf R})$, which is simply an exponential rescaling of $V({\bf x}; {\bf R})$, will again be unable to describe the status of the vacancy. {The high temperature MD test described in Sec.~\ref{sec:ResultsAndDiscussion-AnalysiOfTheDensity} provides and example of this statement. In Fig.~\ref{fig:classicalVsQuantumDensity} we compare the classical and quantum densities at the configuration of the third panel of Fig.~\ref{fig:nonAcceleratedVacancy-Density}, roughly corresponding to the mid-way configuration along this local migration event. The classical density was computed at ${\bar \beta} = 0.1$, the lowest ${\bar \beta}$ at which its value on crystal minima is negligible. The quantum density is bimodal, with one mode localized on each of the crystal sites involved in the migration path. On the hand, the classical density is trimodal, with one mode on one crystal site and two modes on the other. This test shows that already for this simple system, when the temperature is finite, the distribution $\rho_C({\bf x}| {\bf R})$ is inadequate to represent the state of a vacancy.}

\begin{figure}
\begin{center}
\includegraphics[width=0.5\textwidth]{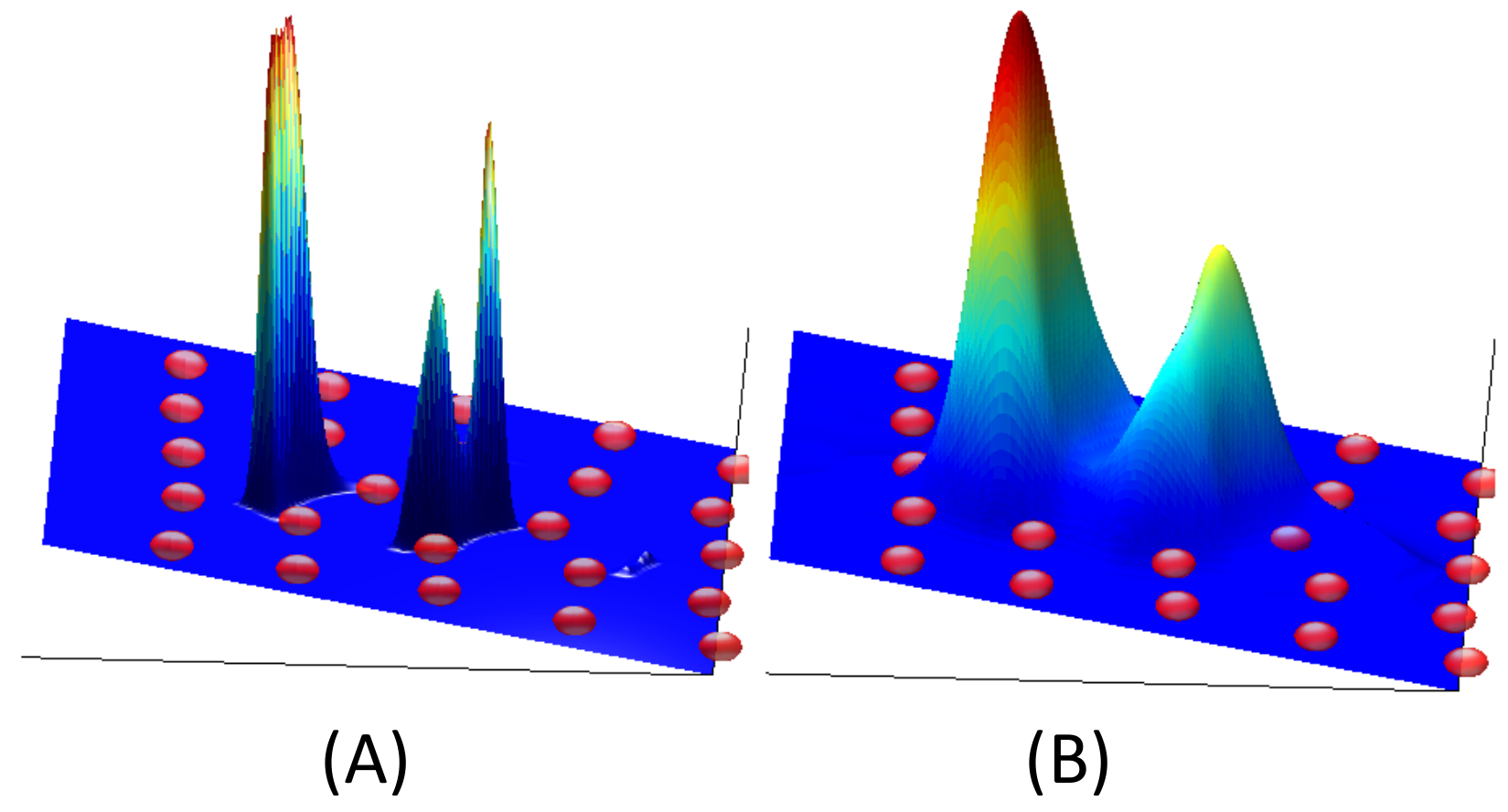}
\end{center}
\caption{{3D representation of the classical (A) and quantum (B) densities in finite temperature system at the same configuration of the third panel of Fig.~\ref{fig:nonAcceleratedVacancy-Density}. The classical density has been computed at $\beta = 0.1$, the lowest $\beta$ with as associated density with negligible density on crystal minima. The scales of the panels (A) and (B) is set such that their maxima have a comparable magnitude.}}\label{fig:classicalVsQuantumDensity}
\end{figure}

\section{The case of ab initio force fields.}
\label{App:QuantumPotentialCase}

Perhaps the most complex extension one can imagine of our representation of a single vacancy is  to the case of a system interacting {\itshape via} an {\itshape ab initio} force field. In this case, our proposal is to represent the vacancy in terms of a probe ``atom'', with its nucleus and electrons equal in charge and number, respectively, to that of the missing atom. Since we assume that the Born-Oppenheimer approximation holds, the electrons will not appear explicitly in the representation of the vacancy, which will be characterized {\itshape via} the nucleus of the probe atom. The electrons of the probe atom will nevertheless contribute to the characterization of the vacancy as they determine the effective potential energy of the interaction of the nucleus of the probe atom with the nuclei of the real atoms of the sample. Consistently with this, and the notation used in the rest of the manuscript, in the following we will denote the positions of the nuclei of the atoms and of the probe atom with ${\bf R}$ and ${\bf x}$, respectively. Let us now consider the system composed of the nuclei of the physical atoms and of the probe particle, and the corresponding electrons in their ground state. Within the Born-Oppenheimer approximation the nuclei interact {\itshape via} the potential energy $E({\bf x}, {\bf R}) = \langle {\mathcal H}({\bf r}; {\bf x}, {\bf R}) \rangle_{gs} $, where ${\bf r}$ denotes the positions of the real electrons plus the electrons of the probe particle, ${\mathcal H}({\bf r}; {\bf x}, {\bf R})$ is the Hamiltonian of the system, and $\langle \cdot \rangle_{gs}$ denotes the expectation value taken over the ground state wavefunction of ${\mathcal H}({\bf r}; {\bf x}, {\bf R})$. We identify the potential $V({\bf x}; {\bf R})$ of Eq.~\ref{eq:potential} with $E({\bf x}, {\bf R})$. The rest of the vacancy representation is unaffected by the fact that the atoms interact {\itshape via} an {\itshape ab initio} force field.

%

\end{document}